\documentstyle[12pt,graphicx]{article}
\begin{document}
\title{Vacuum Polarization Effects in Higher Dimensional Global Monopole Spacetime}
\author{E. R. Bezerra de Mello \thanks{E-mail: emello@fisica.ufpb.br}\\
Departamento de F\'{\i}sica-CCEN, Universidade Federal da Para\'{\i}ba\\
58.059-970, J. Pessoa, PB, C. Postal 5.008, Brazil}
\maketitle

\begin{abstract} In this paper we analyse the vacuum polarization effects associated with a massless scalar field in higher-dimensional global monopole spacetime. Specifically we calculate the renormalized vacuum expectation value of the field square, $\langle\Phi^2(x)\rangle_{Ren}$, induced by a global monopole. Two different spacetimes will be considered: $i)$ In the first, the global monopole lives in whole universe, and $ii)$ in the second, the global monopole lives in a $n=3$ dimensional sub-manifold of the higher-dimensional (bulk) spacetime in the "braneworld" scenario. In order to develop these analysis we calculate the general Euclidean scalar Green function for both spacetimes. Also a general curvature coupling parameter between the field and the geometry is admitted. We explicitly show that $\langle\Phi^2(x)\rangle_{Ren}$ depends crucially on the dimension of the spacetime and on the specific geometry adopted to describe the world. We also investigate the general structure of the renormalized vacuum expectation value of the energy-momentum tensor, $\langle T_{\mu\nu}(x)\rangle_{Ren.}$.
\end{abstract}

\renewcommand{\thesection}{\arabic{section}.}
\section{Introduction}

Different types of topological defects \cite{VS} may have been formed during the phase transitions in the early Universe. Depending on the topology of the vacuum manifold these are domain walls, strings, monopoles and textures. Physically these topological defects appear as a consequence of spontaneous breakdown of local or global gauge symmetries of the system. Global monopoles are spherically symmetric topological defects created due to phase transition when a global symmetry of a system is spontaneously broken. 

The simplified global monopole has been introduced by Sokolov and Starobinsky \cite{SS}. Barriola and Vilenkin \cite{BV} have determined the gravitational field produced by a global monopole in a four-dimensional spacetime, considering a system comprising by a self-coupling iso-scalar Goldstone field triplet $\phi^a$, whose original global $O(3)$ symmetry is spontaneously broken to $U(1)$. The matter field plays the role of an order parameter which outside the monopole's core acquires a non-vanishing value. The main part of the monopole's energy is concentrated into its small core. Coupling this system with the Einstein equations, a spherically symmetric metric tensor is found. Neglecting the small size of the monopole's core, this tensor can be approximately given by the line element
\begin{equation}
ds^{2}=-dt^{2}+\frac{dr^{2}}{\alpha^2}+r^{2}(d\theta^{2}+\sin ^{2}\theta d\phi^{2}) \ ,
\end{equation}
where the parameter $\alpha^2$, smaller than unity, depends on the symmetry breaking energy scale. 

Similarly to a gauge cosmic string \cite{V,G}, a global monopole exerts essentially no gravitational interaction on the surrounding matter; however Barriola and Vilenkin noticed that it acts as a gravitational lens in the same manner as a cosmic string. So, this object may have important role in the cosmology and astrophysics.

Although topological defects have been first analysed in four-dimensional spacetime \cite{VS}, they have been considered in the context of braneworld. In this scenario the topological defects live in a $n-$dimensions submanifold embedded in a $D=4+n$ dimensional Universe. The domain wall case, with a single extra dimension, has been considered in \cite{Rubakov}. More recently the cosmic string case, with two additional extra dimensions, has been analysed in \cite{Cohen,Ruth}. For the case with three extra dimensions, the 't Hooft-Polyakov magnetic monopole has been numerically analysed in \cite{Roessl,Cho}. In Refs. \cite{Ola,Tony,Ola1,Cho1,Cho2} numerical analysis of global monopole are presented.

The calculation of the vacuum polarization effects due to four-dimensional global monopole on the scalar and fermionic fields, have been developed in \cite{ML} and \cite{Mello}, respectively. Here we shall analyse this effect on a quantum massless scalar field considering that the dimension of the spacetime is greater than four. In this way, two distinct topological spacetimes will be considered:\\
$a)$ In the first, the global monopole lives in the whole $D=1+d$ dimensional Universe. In this case the metric tensor associated with this spacetime can be given by the following line element
\begin{equation}
\label{g1}
ds^2_{(a)}=-d t^2+\frac{dr^2}{\alpha^2}+r^2d\Omega_{d-1}^2=g_{MN}d^M dx^N \ ,
\end{equation}
where $M,\  N=0, 1, 2 ...    d$, with $d\geq3$ and $x^M=(t, r, \theta_1, \theta_2, ...  ,\theta_{d-2}, \phi)$. The coordinates are defined in the intervals $t\in(-\infty, \infty)$, $\theta_i\in [0, \pi]$ for $i=1, 2 ... d-2$, $\phi\in[0, 2\pi]$ and $r\geq0$. In this coordinate system the metric tensor is explicitly defined as shown below:
\begin{eqnarray}
\label{Metric}
g_{00}=-1 \ , \ g_{11}=1/\alpha^2 \ , \ g_{22}=r^2 \ \mbox{and} \ g_{jj}=r^2\sin^2\theta_1\sin^2\theta_2 ...\sin^2\theta_{j-2} \ ,
\end{eqnarray}
for $3\leq j \leq d$, and $g_{MN}=0$ for $M\neq N$. This spacetime corresponds to a pointlike global monopole. It is not
flat: the scalar curvature is given by $R=(d-1)(d-2)(1-\alpha^2)/r^2$, and the solid angle associated with a hypersphere with unity radius is $\Omega=2\pi^{d/2}
\alpha^2/\Gamma(d/2)$, so smaller than ordinary one.\\
$b)$ In the second, the global monopole lives in a three dimensional sub-manifold of higher dimensional (bulk) spacetime, having its core in our Universe described by a transverse flat $(p-1)-$dimensional brane. In this case the metric tensor associated with this spacetime is
\begin{eqnarray}
\label{g2}
	ds^2_{(b)}=\eta_{\mu\nu}dx^\mu dx^\nu+\frac{dr^2}{\alpha^2}+r^2d\Omega_2^2=g_{MN}d^M dx^N  \ ,
\end{eqnarray}
where $\eta_{\mu\nu}=diag(-1 \  , \ 1 \ , \ ... \ , \ 1)$ is the Minkowski metric. The curvature scalar associated with this manifold is $R=2(1-\alpha^2)/r^2$, and the solid angle associated with a sphere of unity radius is $\Omega=4\pi\alpha^2$.\\

\section{Euclidean Scalar Green Function}

In order to develop the analysis of the vacuum polarization effects associated with a scalar field, one of the most important quantity is its Green function. Here, in this section, we shall calculate this function admitting that the matter field propagates in the whole space.

The Euclidean Green function associated with a massless scalar field can be obtained by solving the non-homogeneous second order differential equation
\begin{equation}
\label{B}
\left(\Box-\xi R\right)G_E(x,x')=-\delta^D(x,x')=-\frac{\delta^D(x-x')}{\sqrt{g}} \ ,
\end{equation}
with
\begin{eqnarray}
	\Box=\frac1{\sqrt{g}}\partial_M[{\sqrt{g}}g^{MN}\partial_N] \ .
\end{eqnarray}
We have performed in the metric tensors defined by (\ref{g1}) and (\ref{g2}) a Wick rotation $t=i\tau$ on the temporal coordinates. Moreover we have introduced in (\ref{B}) an arbitrary curvature coupling $\xi$.

The Euclidean Green function can also be obtained by the Schwinger-DeWitt formalism as follows:
\begin{equation}
\label{Heat}
G_E(x,x')=\int^\infty_0 ds K(x,x';s) \ ,
\end{equation}
where the heat kernel, $K(x,x';s)$, can be expressed in terms of a complete set of normalized eigenfunctions of the operator $\Box-\xi R$ as follows:
\begin{equation}
\label{Heat-1}
K(x,x';s)=\sum_\sigma \Phi_\sigma(x)\Phi_\sigma^*(x') \exp(-s\sigma^2) \ ,
\end{equation}
with $\sigma^2$ being the corresponding positively defined eigenvalue. Writing
\begin{equation}
\left(\Box-\xi R\right)\Phi_\sigma(x)=-\sigma^2 \Phi_\sigma(x) \ ,
\end{equation}
we obtain the complete set of normalized solutions of the above equation:\noindent\\
For the metric spacetime defined by (\ref{g1}), we have \cite{Mello1}, 
\begin{equation}
\label{PS1}
\Phi_\sigma(x)=\sqrt{\frac{\alpha p}{2\pi}}\frac1{r^{d/2-1}}e^{-i\omega\tau}J_{\nu_l}(pr)Y(l, m_j; \phi, \theta_j) \ ,
\end{equation}
$Y(l,m_j;\phi, \theta_j)$ being the hyperspherical harmonics of degree $l$ \cite{Y}, and $J_{\nu_l}$ the Bessel function of order
\begin{eqnarray}
	\nu_l= \alpha^{-1}\sqrt{\left(l+(d-2)/2\right)^2+(d-1)(d-2)(1-\alpha^2)
(\xi-\overline{\xi})} \ ,
\end{eqnarray}
with the conformal coupling $\overline{\xi}=\frac{d-2}{4(d-1)}$.\\
For metric spacetime defined by (\ref{g2}), we have \cite{Mello2},
\begin{equation}
\label{PS2}
\Phi_\sigma(x)=\frac{{\sqrt{\alpha p}}e^{-ikx}J_{\nu_l}(pr)Y_{lm}(\theta, \ \phi)}{(2\pi)^{p/2}{\sqrt{r}}} \ ,
\end{equation}
with 
\begin{equation}
\nu_l=\alpha^{-1}\sqrt{\left(l+1/2\right)^2+2(1-\alpha^2)(\xi-1/8)} \ ,
\end{equation}
$kx=\eta_{\mu\nu}k^\mu k^\nu $, and $Y_{lm}(\theta,\phi)$ the ordinary spherical harmonics.

In (\ref{PS1}) $\sigma^2=\omega^2+\alpha^2p^2$, and in (\ref{PS2}) $\sigma^2=k^2+\alpha^2p^2$.

Substituting the above expressions in the definition of the heat kernel (\ref{Heat-1}) and using (\ref{Heat}), we obtain the following Green functions:\\
For the spacetime defined by the metric tensor (\ref{g1}),
\begin{eqnarray}
\label{Green-a}
G^{(a)}_E(x,x')=\frac1{4\pi^{d/2+1}}\frac1{(rr')^{\frac{d-1}2}}\frac{\Gamma(d/2)}{d-2}
\sum_{l=0}^\infty[2(l-1)+d]Q_{\nu_l-1/2}(\cosh u_a)C_l^{\frac{d-2}2}(\cos\gamma) \ ,
\end{eqnarray}
where
\begin{equation}
\cosh u_a=\frac{\alpha^2\Delta\tau^2+r^2+r'^2}{2rr'} \ ,
\end{equation}
and $C_l^\mu(x)$ being Gegenbauer polynomial of degree $l$.\\
As to the spacetime defined by (\ref{g2}), the Green function reads
\begin{eqnarray}
\label{Green-b}
G^{(b)}_E(x,x')=\frac1{2^{\frac{p+5}2}\pi^{\frac{p+3}2}}\frac{i^{1-p}}{\alpha^{1-p}}\frac1{(rr')^{\frac{p+1}2}}\frac1{(\sinh u_b)^{\frac{p-1}2}}\sum_{l=0}^\infty(2l+1)Q_{\nu_l-1/2}^{\frac{p-1}2}(\cosh u_b)P_l(\cos\gamma) \ ,
\end{eqnarray}
with
\begin{equation}
\label{u}
\cosh u_b=\frac{\Delta x^2\alpha^2+r^2+r'^2}{2rr'} \ .
\end{equation} 

In both Green functions $(Q^\lambda_\nu)$ $Q_\nu$ is the (associated) Legendre function, and $\gamma$ the angle between the two arbitrary directions.

\section{The Computation of $\langle\Phi^2(x)\rangle_{Ren.}$}
The vacuum expectation value of the square of the scalar field is formally expressed by taking the coincidence limit of the Green function as shown below:
\begin{equation}
\langle\Phi^2(x)\rangle=\lim_{x'\to x}G_E(x,x') \ .
\end{equation}
However this procedure provides a divergent result. In order to obtain a finite and well defined result, we must apply some renormalization procedure. Here we shall adopt the point-splitting renormalization one. The basic idea of this procedure is to analyse the divergent contributions of the Green function in the coincidence limit and subtract them off. In \cite{Wald}, Wald observed that the singular behavior of the Green function has the same structure as given by the Hadamard one, which on the other hand can be written in terms of the square of the geodesic distance between two points. So, here we shall adopt the following prescription: we subtract from the Green function the Hadamard one before applying the coincidence. In this way, the renormalized vacuum expectation value of the field square is given by:
\begin{equation}
\label{Phi2}
\langle\Phi^2(x)\rangle_{Ren.}=\lim_{x'\to x}\left[G_E(x,x')-G_H(x,x')\right] \ .
\end{equation}
Because the explicit expression of the Hadamard function depends on the dimension of the spacetime, the above calculation can only be explicitly performed by specifying the dimensions of the spacetime. So in the next sub-sections we shall consider spacetimes with $5$ and $6$ dimensions.

\subsection{Five Dimensional Spacetime}
In order to develop the analysis for $\langle\Phi^2(x)\rangle$, it is necessary to write explicitly the Green functions taking $d=4$ in (\ref{Green-a}) and $p=2$ in (\ref{Green-b}); moreover, it is also necessary to adopt for the respective Legendre functions specific representations. Here we shall adopt integral representations as follows:\\
For the first function, we shall use \cite{Grad}:
\begin{equation}
\label{Legendre}
Q_{\nu_l-1/2}(\cosh u_a)=\frac1{\sqrt{2}}\int^\infty_{u_a} dt \frac{e^{-\nu_l t}}{\sqrt{\cosh t-\cosh u_a}}
\end{equation}
and for the second, we shall use \cite{Grad}:
\begin{eqnarray}
	Q_{\nu_l-1/2}^{1/2}(\cosh u_b)=i{\sqrt\frac\pi2}\frac{e^{-\nu_lu_b}}{{\sqrt{\sinh u_b}}} \ .
\end{eqnarray}

Because the orders of the Legendre functions, $\nu_l$, depend on the parameter $\alpha$ in a very complicate form, it is not possible to proceed exactly their respective summation on the quantum number $l$ in (\ref{Green-a}) and (\ref{Green-b}). The best we can do, is to develop a series expansion in powers of the parameter $\eta^2=1-\alpha^2$ considered much smaller than unity\footnote{In fact, for a typical grand unified theory in four dimensions, the parameter $\eta^2$ is of order $10^{-5}$}. The expansions are:
\begin{equation}
\nu_l\approx (l+1)(1+\eta^2/2)+\frac{(3\xi-1/2)}{l+1}\eta^2+O(\eta^4) \ ,
\end{equation}
for the case $(a)$, and 
\begin{eqnarray}
\nu_l\approx(l+1/2)(1+\eta^2/2)+\frac{(2\xi-1/4)}{2l+1}\eta^2 +O(\eta^4) \ ,	
\end{eqnarray}
for the case $(b)$.

Taking first the coincidence limit in the angular variables, and after some intermediates steps, the Euclidean Green functions read:\\
For the case $(a)$:
\begin{eqnarray}
\label{Ga5}
G^{(a)}(r,r')&=&\frac1{16\sqrt{2}\pi^3}\frac1{(rr')^{3/2}}\int^\infty_{u_a}
dt \frac1{\sqrt{\cosh t-\cosh u_a}}\frac{\cosh(t/2)}{\sinh^3(t/2)}\times\nonumber\\
&&\left[1-\frac{3t\eta^2}{2\sinh(t)}(1+4\xi\sinh^2(t/2))\right]+O(\eta^4) \ .
\end{eqnarray}
For the case $(b)$:
\begin{eqnarray}
\label{Gb5}
	G^{(b)}(x,x')=\frac1{64\pi^2}\frac1{(rr')^{3/2}}\frac{(1+\eta^2)}{\sinh^3u_b}\left[1-\frac{u_b\eta^2}{\sinh u_b}\left(1+ 4\xi\sinh^2(u_b/2)\right)\right]+O(\eta^4) \ .
\end{eqnarray}

The general expression to the Hadamard function for scalar fields in the spacetime of odd dimensions has been given in \cite{Chris}. For a five-dimensional spacetime the Hadamard function reads,
\begin{equation}
\label{H5}
G_H(x,x')=\frac1{16\sqrt{2}\pi^2}\frac1{\sigma^{3/2}(x,x')}\left[1+(1/6-\xi)R(x)\sigma(x,x')\right] \ ,
\end{equation}
being $\sigma(x',x)$ the one-half of the square of the geodesic distance between two arbitrary points, and $R$ the scalar curvature. The one-half of the radial geodesic distances for both spacetimes read $\sigma(x,x')=(1/2\alpha^2)(r-r')^2$. In our approximation it can be written as $\sigma\approx(1/2)(r-r')^2(1+\eta^2+...)$.

Now we are in position to calculate the renormalized vacuum expectation value of the square of the field operator up to the first order in $\eta^2$. Once more the two distinct situations have to be analysed separately:\\
$i)$ For the spacetime defined by (\ref{g1}) the scalar curvature is $R=6\eta^2/r^2$. Substituting (\ref{Ga5}) and (\ref{H5}) into (\ref{Phi2}) we get
\begin{equation}
\langle\Phi^2(x)\rangle_{Ren.}=\frac{3\eta^2}{64\pi r^3}\left(\xi-3/16\right) \ .
\end{equation}
We can see that for the conformal coupling in five dimensional spacetime, $\xi=3/16$, the renormalized vacuum expectation value of the operator $\Phi^2(x)$ is zero, up to the first order in $\eta^2$.\\
$ii)$ For the spacetime defined by (\ref{g2}) the scalar curvature is $R=2\eta^2/r^2$. Substituting (\ref{Gb5}) and (\ref{H5}) into (\ref{Phi2}) we get a vanishing result:
\begin{eqnarray}
\langle\Phi^2(x)\rangle_{Ren.}=0 	
\end{eqnarray}
for any value of the non-minimal coupling constant $\xi$.

Because the above vanishing result, we may want to know the vacuum expectation of the field square in the next-to-leading order, i.e., at order $O(\eta^4)$. To do that, we have to construct the Green and Hadamard functions up to this order. Developing a long calculation \cite{Mello2}, we finally get a non-vanishing result:
\begin{eqnarray}
	\langle\Phi^2(x)\rangle_{Ren.}=-\frac{\eta^4}{192r^3}\left(\xi-1/8\right) \ . 
\end{eqnarray}

\subsection{Six Dimensional Spacetime}
Following the same steps, the Euclidean Green function for the spacetime defined by (\ref{g1}), in six dimensions (d=$5$) reads
\begin{eqnarray}
\label{Ga}
G^{(a)}(r,r')&=&\frac3{128\sqrt{2}\pi^3}\frac1{(rr')^{2}}\int^\infty_{u_a}
dt \frac1{\sqrt{\cosh t-\cosh u_a}}\frac{\cosh(t/2)}{\sinh^4(t/2)}\times\nonumber\\
&&\left[1-\frac{2t\eta^2}{\sinh(t)}(1+4\xi\sinh^2(t/2))\right] \ .
\end{eqnarray}
As to the spacetime defined by (\ref{g2}), the Green function (p=$3$) reads
\begin{eqnarray}
\label{Gb}
G^{(b)}(x,x')=-\frac{1-\eta^2}{2^4\pi^3}\frac1{(rr')^2}\frac1{\sinh u_b}\sum_{l=0}^\infty(2l+1)Q_{\nu_l-1/2}^1(\cosh u_b)P_l(\cos\gamma) \ .
\end{eqnarray}
In order to investigate the he vacuum polarization effect we shall use in (\ref{Gb}) the integral representation to the associated Legendre function given below \cite{Grad}:
\begin{eqnarray}
\label{Q1}
Q_{\nu-1/2}^\lambda(\cosh u)={\sqrt{\frac\pi 2}}\frac{e^{i\lambda\pi}\sinh^\lambda(u)}{\Gamma(1/2-\lambda)}\int_u^\infty \ dt \ \frac{e^{-\nu t}}{(\cosh t- \cosh u)^{\lambda+1/2}} \ .
\end{eqnarray}
For this case $\lambda=(p-1)/2=1$. However the above representation can only be applied for $Re(\lambda)<1/2$.  This integral representation, on the other hand, can be used for submanifold $(p-1)-$brane of smaller dimension. In the calculation of vacuum polarization effects, we have adopted the point-splitting renormalization procedure, subtracting from the Green function the Hadamard one. This procedure provides a finite and well defined result to evaluate the renormalized vacuum expectation value of the square of the scalar field. In what follows, we shall allow in this renormalization procedure, that the dimension of the brane be an arbitrary number. In this way we may use (\ref{Q1}) in Green function above, and also in the definition of Hadamard function. Finally, in the calculation of the vacuum polarization effect, we shall take $p\to 3$ before to take the coincidence limit in the renormalized Green function. As we shall see we shall obtain a finite and well defined result. Adopting this procedure the Green function can be written by
\begin{eqnarray}
\label{Gr3}
	G^{(b)}(x',x)&=&\frac{{\sqrt{2}}}{32\pi^2{\sqrt{\pi}}}\frac{\alpha^2}{(rr')^2}\frac1{\Gamma(1/2-\lambda)}\int_u^\infty\frac{dt} {(\cosh t-\cosh u)^{\lambda+1/2}}\times\nonumber\\
&&\sum_{l=0}^\infty(2l+1)e^{-\nu_lt}P_l(\cos\gamma) \ .
\end{eqnarray}
Taking $\gamma=0$ ($P_l(1)=1$) into the above equation it is possible to develop an approximated expression to the summation on the angular quantum number $l$.

The Hadamard function in a six dimensional spacetime has the general form:
\begin{eqnarray}
\label{H2}		G_H(x',x)=\frac{\Delta^{1/2}(x,x')}{16\pi^3}\left[\frac{a_0(x,x')}{\sigma^2(x,x')}+\frac{a_1(x,x')}{2\sigma(x,x')}-\frac{a_2(x,x')}4\ln\left(\frac{\mu^2\sigma(x,x')}2\right)\right] \ ,
\end{eqnarray}
where $\mu$ is an arbitrary energy scale introduced in this formalism to prevent infrared singularity, $\Delta(x,x')$ is the Van Vleck-Morette determinant and the coefficients, $a_k(x,x')$, for $k=0, \ 1,\ 2$, have been computed by many authors\footnote{See Refs. \cite{BS} and \cite{Chr1}.}. For the radial point-splitting we have $\sigma(x',x)= (r'-r)^2/2\alpha^2$. 

The expressions for the coefficients $a_k$ depend on the scalar curvature, Ricci tensor, etc. For the metric tensor defined by (\ref{g1}), the Hadamard function, up to the first order expansion in the parameter $\eta^2$, reads:
\begin{equation}
\label{Ha}
G_H^{(a)}(r,r')=\frac1{16\pi^3}\left[\frac{4(1-2\eta^2)}{(r-r')^4}+\frac{2(1-6\xi)\eta^2}{r^2(r-r')^2}-\frac{\eta^2}{r^4}(\xi-1/5)
\ln\left(\mu^2(r-r')^2/4\right)\right] \ .
\end{equation}
However for the metric tensor defined by (\ref{g2}) we have:
\begin{eqnarray}
\label{Hb}
G_H^{(b)}(r',r)=\frac1{16\pi^3}\left[\frac{4(1-2\eta^2)}{(r-r')^4}+\frac{(1-6\xi)\eta^2}{3r^2(r-r')^2}
-\frac{\eta^2}{6r^4}\left(1/5-\xi\right)\ln\left(\mu^2(r-r')^2/4\right)\right]	\ .
\end{eqnarray}

At this point we shall adopt the approach below to express the Hadamard functions in a integral representation. We shall express the different powers of $\frac1{r'-r}$ in the Hadamard functions above by the following integral representation:\\
For the Hadamard function defined in (\ref{Ha}), we use
\begin{eqnarray}
	\frac1{(r'-r)^{d-1}}&=&\frac{\sqrt{2}\ \Gamma(\frac d2)}{2^{d-1}(rr')^{\frac{d-1}{2}}\sqrt{\pi}\Gamma(\frac{d-1}2)}\times\nonumber\\
&&	\int_u^\infty\frac{dt}{\sqrt{\cosh t-\cosh u}} \frac{\cosh(t/2)}{\sinh^{d-1}(t/2)} \ ,
\end{eqnarray}
and for the Hadamard function (\ref{Hb}), 
\begin{eqnarray}
	\frac1{(r'-r)^{d+1}}&=&\frac{(r'-r)^{2(\lambda-1)}}{2^{d+\lambda-\frac32}}\frac1{(r'r)^{\frac{d+2\lambda-1}2}}\frac{\Gamma(\frac d2)}{\Gamma(\frac{d-1}2+\lambda)\Gamma(\frac12-\lambda)}\times\nonumber\\
&&	\int_u^\infty\frac{dt}{(\cosh t-\cosh u)^{\frac12+\lambda}} \frac{\cosh(t/2)}{\sinh^{d-1}(t/2)} \ .
\end{eqnarray}

Substituting the parameter $d$ for the appropriated values in order to reproduce the correct powers of $\frac1{r'-r}$, and expressing the logarithmic term in both functions by $Q_0(\cosh u)$, we obtain two long expressions. The renormalized vacuum expectation values of field square have to be evaluated separately, for both cases:
\noindent\\
$i)$ For the first case we have to substitute (\ref{Ga}) and (\ref{Ha}) into (\ref{Phi2}). Taking the coincidence limit we have
\begin{equation}
\langle\Phi^2(x)\rangle_{Ren.}=-\frac{\eta^2}{96\pi^3r^4}\left(\frac{47}{25}-10\xi\right)+\frac{\eta^2}{8\pi^3r^4}\left(\xi-1/5\right)\ln(\mu r) \ .
\end{equation} 
$ii)$ For the second case we have to substitute (\ref{Gr3}) and (\ref{Hb}) into (\ref{Phi2}). However, as we have mentioned before, we shall take $\lambda\to1$ first into the renormalized Green function before to take the coincidence limit. Doing this procedure we get:
\begin{equation}
\langle\Phi^2(x)\rangle_{Ren.}=\frac1{576\pi^3}\frac{\eta^2}{r^4}\left(\frac{47}{25}-10\xi\right)+\frac1{48\pi^3}\frac{\eta^2}{r^4}\left(\xi-1/5\right)\ln(\mu r) \ .
\end{equation}

We can see that, although both results above are different, there are some similarities between them:\\
$a)$ For the conformal coupling in six dimension, $\xi=1/5$, there is no ambiguity in the definition of the above vacuum polarization effects, i.e., the logarithmic contributions disappear, and\\
$b)$ for $\xi=47/250$ the contributions proportional to $1/r^4$ disappear.

\section{Energy-Momentum Tensor}

In this work we are analyzing the quantum effects associated with a massless scalar field in the metric spacetimes defined by (\ref{g1}) and (\ref{g2}). As we can see these metric tensors present no dimensional parameter. Moreover we are adopting the natural system units where $\hbar=c=1$. As a consequence we can conclude that any physical quantities calculated can only depend on the radial coordinate $r$ or on the renormalization mass scale $\mu$. By dimensional arguments we could expect that $\langle\Phi^2(x)\rangle_{Ren.}$ is proportional to $1/r^{n-2}$ and  $\langle T_{MN}(x)\rangle_{Ren.}$ proportional to $1/r^n$, being $n$ the dimension of the spacetime considered. The factor of proportionality should been given in terms of the parameter $\eta^2$ and the non-minimal coupling $\xi$. In this section we want to analyse the renormalized vacuum expectation value (VEV) of the energy-momentum tensor. By calculations developed previously, we have shown that, up to the first order in $\eta^2$, the renormalized VEV of the field square in the metric defined by (\ref{g2}) is zero for a five-dimensional spacetime\footnote{For the spacetime defined by (\ref{g1}) we have seen that this vacuum expectation value does not vanish.}. Although we cannot affirm that these vanishing result also occur in the calculation of the VEV of the energy-momentum tensor, we shall analyse $\langle T_{MN}(x)\rangle_{Ren.}$ for the six-dimensional spacetime only. 

The renormalized vacuum expectation value of the energy-momentum tensor should obey the conservation condition
\begin{equation}
\label{Con}
\nabla_M\langle T_N^M(x)\rangle_{Ren.}=0 \ ,
\end{equation}
and provides the correct trace anomaly. For a six-dimensional spacetime it reads \cite{Chr2}:
\begin{equation}
\label{Ano}
\langle T_M^M(x)\rangle_{Ren.}=\frac1{64\pi^3}a_3(x) \ .
\end{equation}

Taking into account all above informations, we can conclude that the general structure for the renormalized vacuum expectation value of the energy-momentum is:
\begin{equation}
\langle T_M^N(x)\rangle_{Ren.}=\frac1{64\pi^3r^6}\left[A_M^N(\eta^2,\xi)+B_M^N(\eta^2,\xi)\ln(\mu r)\right] \ ,
\end{equation}
with $A_M^N$ obeying specific restriction conditions that will be examined later. Because the cutoff factor $\mu$ is completely arbitrary, there is an ambiguity in the definition of this renormalized vacuum expectation value. Moreover the change in this quantity under the change of the renormalization scale is given in terms of the tensor $B_\mu^\nu$ as shown below:
\begin{equation}
\langle T_M^N(x)\rangle_{Ren.}(\mu)-\langle T_M^N(x)\rangle_{Ren.}(\mu')=\frac1{64\pi^3r^6}B_M^N(\eta^2,\xi)\ln(\mu/\mu') \ .
\end{equation}
The difference between them is given in terms of the effective action which depends on the logarithmic terms whose final expression, in arbitrary even dimension, is \cite{Chr2}:
\begin{equation}
\langle T_{MN}(x)\rangle_{Ren.}(\mu)-\langle T_{MN}(x)\rangle_{Ren.}(\mu')=\frac1{(4\pi)^{n/2}}\frac1{\sqrt{g}}\frac\delta{\delta g^{MN}}
\int d^nx\sqrt{g}a_{n/2}(x)\ln(\mu/\mu') \ .
\end{equation}
In our six dimensional case we need the factor $a_3(x)$. The explicit expression for this factor can be found in the paper by Gilkey \cite{Gilkey} and in a more systematic form in the paper by Jack and Parker \cite{JP}, for a scalar second order differential operator $D^2+X$, $D_M$ being the covariant derivative including gauge field and $X$ an arbitrary scalar function. This expression involves $46$ terms and we shall not repeat it here in a complete form. The reason is because our calculation has been developed up to the first order in the parameter $\eta^2$ and only the quadratic terms in Riemann and Ricci tensors, and in the scalar curvature are relevant for us\footnote{The term proportional to $\Box^2 R$ in $a_3(x)$ is also not relevant because it can be written as a total derivative.}. This reduces to $12$ the number of terms which will be considered. Discarding the gauge fields and taking $X=-\xi R$ we get:
\begin{eqnarray}
\overline{a}_3(x)&=&\frac16\left(\frac16-\xi\right)\left(\frac15-\xi\right)
R\Box R+\frac{\xi^2}{12}R^{;M} R_{;M}+\frac\xi{90}R^{MN}
R_{;MN}-\frac{\xi}{36}R^{;M} R_{;M}
\nonumber\\
&&-\frac1{7!}\left[28R\Box R +17R_{;M} R^{;M}-2R_{MN;P}
R^{MN;P}-4R_{MN;P}R^{MP;N}\right.+
\nonumber\\
&&9R_{MNPS;G}
R^{MNPS;G}-8R_{MN}\Box R^{MN}+24R_{MN}
R^{MP;N}\ _P+
\nonumber\\
&&\left.12R_{MNPS}\Box R^{MNPS}
\right]+O(R^3) \ .
\end{eqnarray}
This expression is of sixth order derivative on the metric tensor. Our next step is to take the functional derivative of $\overline{a_3}(x)$. Using the expressions for the functional derivative of the Riemann and Ricci tensor, together with the scalar curvature, we obtain after a long calculation the following expression for the tensor $B_M^N$:
\begin{eqnarray}
B_M^N(\eta^2,\xi)&=&\frac{r^6}6\left[-\delta_M^N\Box^2 R\left(\xi^2-\frac\xi3+\frac{23}{840}\right)+\frac1{140}\Box^2 R_M^N+\right.
\nonumber\\
&&\left.\nabla^N\nabla_M\Box R\left(\xi^2-\frac\xi3+\frac1{42}\right)\right]+O(R^2) \ .
\end{eqnarray}
Developing all the terms which appear in the above equation we obtain after some calculations:\\
$i)$ For the metric spacetime defined by (\ref{g1}):
\begin{eqnarray}
\label{B1}
B_M^N(\eta^2,\xi)&=&\frac{\eta^2}{75}\ diag\ (\ 2, 2, -1, -1, -1, -1\ )+
\nonumber\\
&&16\eta^2(\xi-1/5)(\xi-2/15)\ diag\ (\ 1, -4, 2, 2, 2, 2\ ) \ .
\end{eqnarray}
$ii)$ For the metric spacetime defined by (\ref{g2}):
\begin{eqnarray}
\label{B2}
B_M^N(\eta^2,\xi)&=-&\frac{\eta^2}{175}\ diag\ (\ 1, 1, 1, 1, -2,-2\ )-
\nonumber\\
&&8\eta^2(\xi-1/5)(\xi-2/15)\ diag\ (\ 1, 1, 1, -2/3, 4/3, 4/3\ ) \ .
\end{eqnarray}
We can see that by taking $\xi=1/5$ the trace of both terms vanish.

After conclude the analysis above for the tensor $B_M^N$, let us present below the restriction conditions obeyed by the components of the tensor $A_M^N$, for the spacetimes defined by (\ref{g1}) and (\ref{g2}) separately:\\ 
By applying the conservation condition, (\ref{Con}), and the correct trace anomaly expression, (\ref{Ano}), we can write, for the conform coupling $\xi=1/5$, the following results for spacetime defined by the metric tensor (\ref{g1}):
\begin{eqnarray}
A_1^1&=&A_0^0-T++B_1^1 \nonumber\\
A_2^2&=&A_3^3=A_4^4=A_5^5=\frac{T}2-\frac{A_0^0}2-\frac{B_1^1}4 \ ,
\end{eqnarray}
with
\begin{equation}
T=r^6 a_3(x)=r^6\frac{18}{7!}\Box^2 R +O(R^2)=-\frac{12\eta^2}{35}+O(\eta^4) \ .
\end{equation}
As to the spacetime defined by (\ref{g2}), we observe that the geometry of its brane section has a Minkowski-type structure, consequently we expect that $A_0^0 \ = \ A_1^1= \ A_2^2$. Admitting this fact we can write\footnote{By (\ref{B2}), we can see that $B_0^0=B_1^1=B_2^2$ for any value of curvature coupling $\xi$.}:
\begin{eqnarray}
	A^3_3&=&A^0_0+\frac{B_3^3}3-T \ ,\\
	A_4^4&=&A^5_5=-2A_0^0-\frac{B_3^3}6+\frac{2T}3 \ ,
\end{eqnarray}
with
\begin{eqnarray}
	T=r^6a_3(x)=r^6\frac{18}{7!}\Box^2R+O(\eta^4)=\frac{6\eta^2}{35}+O(\eta^4) \ .
\end{eqnarray}

We conclude this section by saying that the complete evaluation of $\langle T_M^N(x)\rangle_{Ren.}$, for both spacetimes, requires the knowledge of at least one component of the tensor $A_M^N$, for example $A_0^0$. However we shall not attempt to develop this straightforward and long calculation here.

\section{Concluding Remarks}

In this paper we have investigated the vacuum polarization effects associated with a massless scalar field induced by the presence of a global monopole in spacetimes of dimensions higher than four. Two different geometric spacetimes have been considered:\\
\begin{itemize}
\item In the first, the global monopole lives in whole space.
\item In the second, the monopole lives in a three-dimensional submanifold of higher-dimensional (bulk) spacetime.
\end{itemize}

Our main objective in this paper was to investigate how different geometries associated with the same topological object can provide different results at quantum level. In order to answer that question two specific calculations have been developed: the renormalized vacuum expectation values of the field square, $\langle\Phi^2(x) \rangle_{Ren.}$, and the energy-momentum tensor $\langle T_M^N (x)\rangle_{Ren.}$.

As to $\langle\Phi^2(x)\rangle_{Ren.}$, we develop this calculation for spacetimes of five, respectively six dimensions. We have found  that, up to the first order in the parameter $\eta^2=1-\alpha^2$, assumed to be smaller than unity, this quantity presents different results for each geometry considered. In the five dimensional case, the vacuum average gets, in principle, a non-vanishing result for the spacetime defined by (\ref{g1}), and a vanishing result for the spacetime defined by (\ref{g2}).  For the six dimensional one, although being different the values found for $\langle\Phi^2(x)\rangle_{Ren.}$, they present some similarities as mentioned in section $3$.

The renormalized vacuum expectation value of the energy-momentum tensor, has been analyzed for a six dimensional spacetime under dimensional grounds only. We have shown that it behaves as $1/r^6$, where $r$ is the distance from the monopole's core, and presents an additional contribution proportional to $\ln(\mu r)/r^6$, being $\mu$ is an arbitrary mass scale introduced by the renormalization prescription. This term is associated with the coefficient $a_3(x)$, which, according to \cite{BD}, comes from the purely geometric (divergent) Lagrangian that should renormalize the modified classical Einstein one. When this extra term is inserted into the gravitational action, the left-hand side of the field equation is modified by the presence of order six terms proportional to:
\begin{equation}
c_1 g_{AB}\Box^2 R+c_2 \Box^2 R_{AB}+c_3 \nabla_A\nabla_B \Box R +
O(R^2) \ .
\end{equation}

\section*{Acknowledgment} 
The author thanks to Conselho Nacional de Desenvolvimento Cient\'\i fico e Tecnol\'ogico (CNPq.) and FAPESQ-PB/CNPq. (PRONEX) for partial financial support.

\end{document}